\documentclass[vecphys]{svmult}

\usepackage{makeidx}         
\usepackage{graphicx}        
\usepackage{multicol}        
\usepackage[bottom]{footmisc}

\makeindex             

\begin{document}

\title*{The abundance discrepancy problem in H\thinspace{\sc ii} regions}
\titlerunning{The abundance discrepancy problem in H\thinspace{\sc ii} regions}
\author{Jorge Garc\'{\i}a-Rojas\inst{}\and
C\'esar Esteban\inst{}}
\institute{Instituto de Astrof\'{\i}sica de Canarias, c/ Via L\'actea s/n, E-38205, La Laguna, S/C Tenerife, Spain
\texttt{jogarcia@iac.es; cel@iac.es}}
%
%
\maketitle
\index{Author1}
\index{Author2}

\begin{abstract}
The origin of the abundance discrepancy in nebulae is one of the key problems in the physics of photoionized 
nebulae. In this work we have analized a sample of Galactic and extragalactic H\thinspace {\sc ii} regions where  
the abundance discrepancy have been measured, and we discuss the two main scenarios proposed to explain such 
discrepancy: temperature fluctuations over the observed volume of the nebulae and chemically inhomogeneous 
inclusions.
\end{abstract}
 
\section{Overview}
\label{sec:1}

Almost forty years ago, \cite{peimbert67} claimed that in the presence of temperature variations over the observed 
volume the abundances derived for nebulae from the analysis of intensity ratios of colisionally excited lines 
(CELs) were underestimated if these temperature fluctuations were not considered. This is due to the strong 
dependence of the intensity of those lines on the assumed electron temperature. This possibility is puzzling, 
considering that the analysis of CELs is the standard method for deriving ionic abundances in ionized nebulae. On 
the other hand, intensity ratios of recombination lines (RLs) are almost independent of the temperature structure 
of the nebula, and could, in principle, be more appropiate to derive the ``real'' abundances of the nebulae. 
Measurements of RLs on H\thinspace {\sc ii} regions and planetary nebulae (PNe) in the last years have found that 
abundance determinations from RLs are sistematically larger than those obtained using CELs, independently of the 
ion considered (see e.g. \cite{liu06}, and \cite{garciarojasetal06} and references therein).   

\cite{tsamispequignot05} have presented an scenario to explain the abundance discrepancy in H\thinspace {\sc ii} 
regions, which is based on the implicit assumption on the existence of temperature fluctuations produced by 
chemical inhomogeneities in the ionized gas. These authors proposed a model for 30 Doradus, in which they postulate 
the existence of a low temperature, metal-rich gas embedded in an ``ambient'' medium with lower density and 
metallicity and higher temperature. According to \cite{tsamispequignot05}, this denser component comes from Type II 
SN ejecta, that has not been mixed with the interstellar medium and is in pressure equilibrium with the ``normal'' 
chemical composition gas. This droplets would be responsible of most of the RLs emission, and due to their low 
temperature they would not emit CELs. 

In this paper we present results on deep VLT high resolution spectrophotometriy obtained by our group in the last 
years (\cite{estebanetal04, garciarojasetal04, garciarojasetal05, garciarojasetal06, garciarojasetal06b, 
lopezsanchezetal06}) to discuss the validity of the two scenarios described above to explain the abundance 
discrepancy problem. 

\section{Correlations of the ADF with nebular properties}

The different scenarios proposed to explain the abundance discrepancy predict a different behaviour of the ADF with 
respect to different nebular properties. 

\begin{figure}[ht!]
\centering
\includegraphics[width=5.5cm,angle=0]{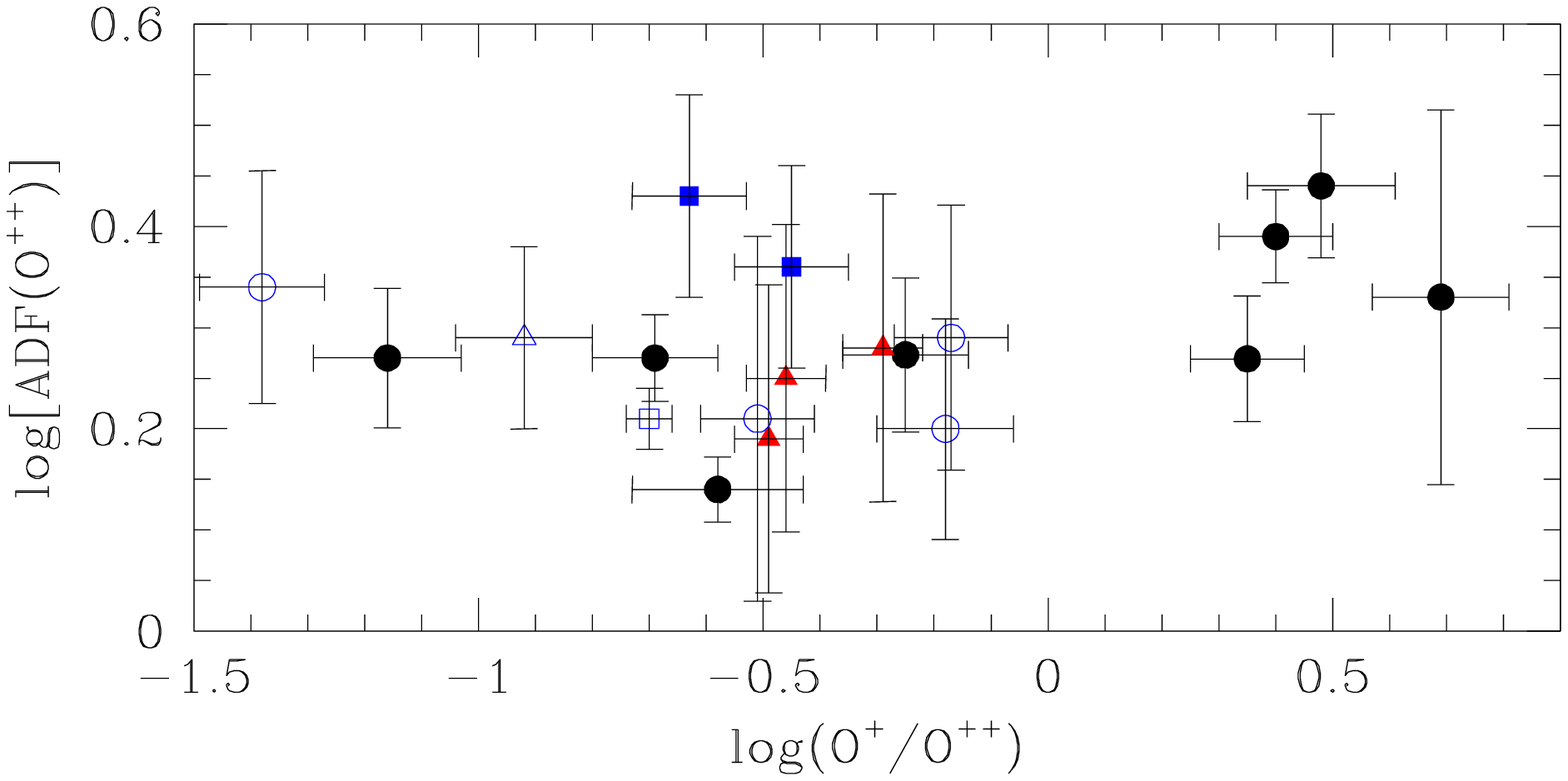}
\includegraphics[width=5.5cm,angle=0]{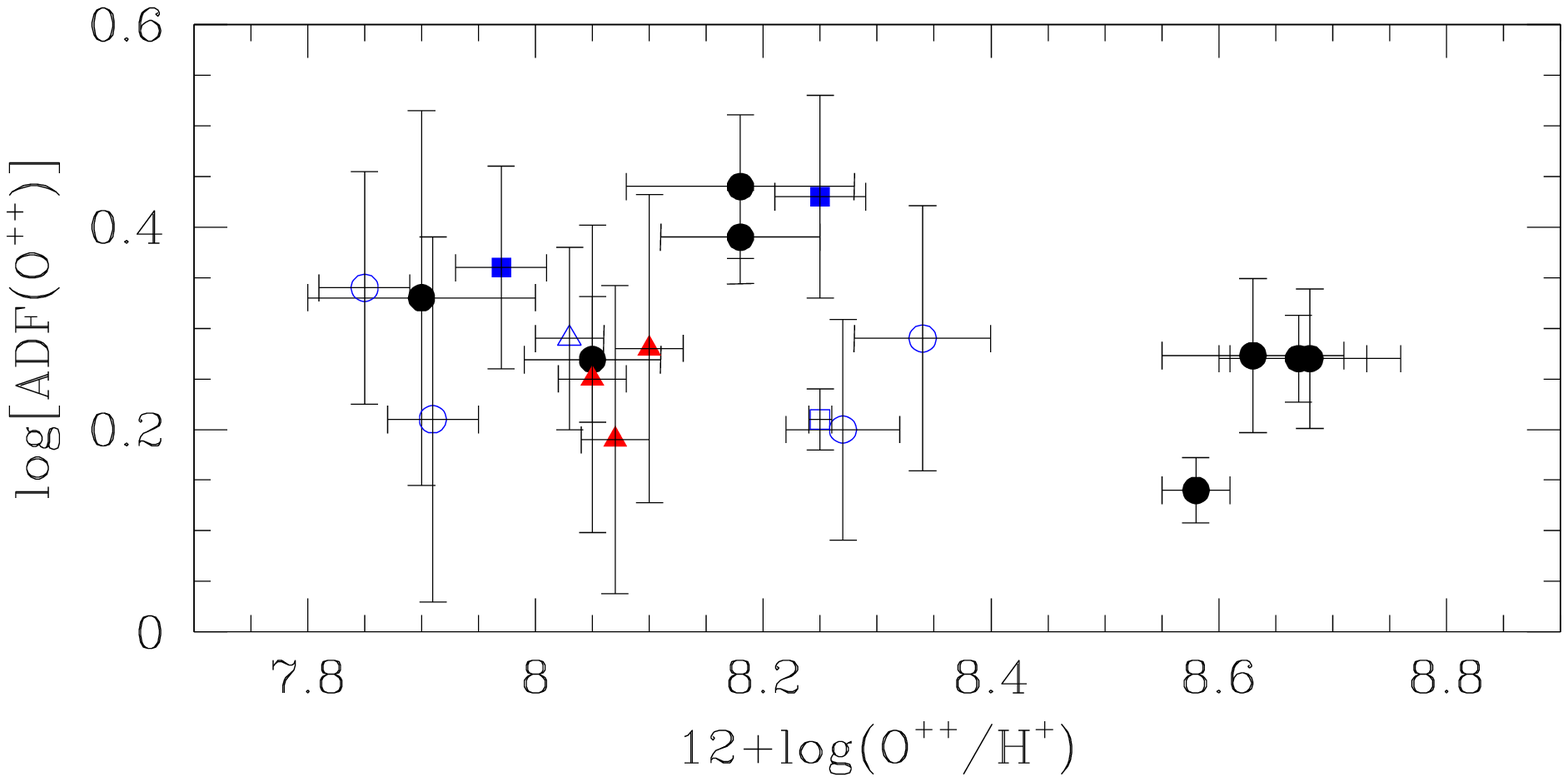}
\protect\caption[ ]{ADF(O$^{++}$) $vs.$ excitation (left) and $vs.$ O$^{++}$/H$^+$ ratio (right). Filled circles: 
Galactic H\thinspace {\sc ii} regions; 
rest of symbols: extragalactic H\thinspace {\sc ii} regions from the literature
(see \cite{lopezsanchezetal06} and references therein).
\label{adfoxy}}
\end{figure}

The metallicity dependence of the $t^2$ parameter was suggested by \cite{garnett92}, who found that photoionization 
models could reproduce temperature fluctuations similar to those observed in the case that electron temperature is  
less than $\sim$9000 K, raising on colder --more metal-rich-- nebulae. This result does not agree with the 
observations in both H\thinspace {\sc ii} regions and PNe. In figure~\ref{adfoxy} we show ADF $vs.$ O$^{++}$/H$^+$ 
and excitation, for all the Galactic and extragalactic  
H\thinspace {\sc ii} regions for which O$^{++}$ abundances from O\thinspace {\sc ii} RLs are available. 
Despite the narrow range of parameters covered, we have found no correlation between ADF(O$^{++}$) and the rest of 
parameters.  

\begin{figure}[ht!]
\centering
\includegraphics[width=5cm,angle=0]{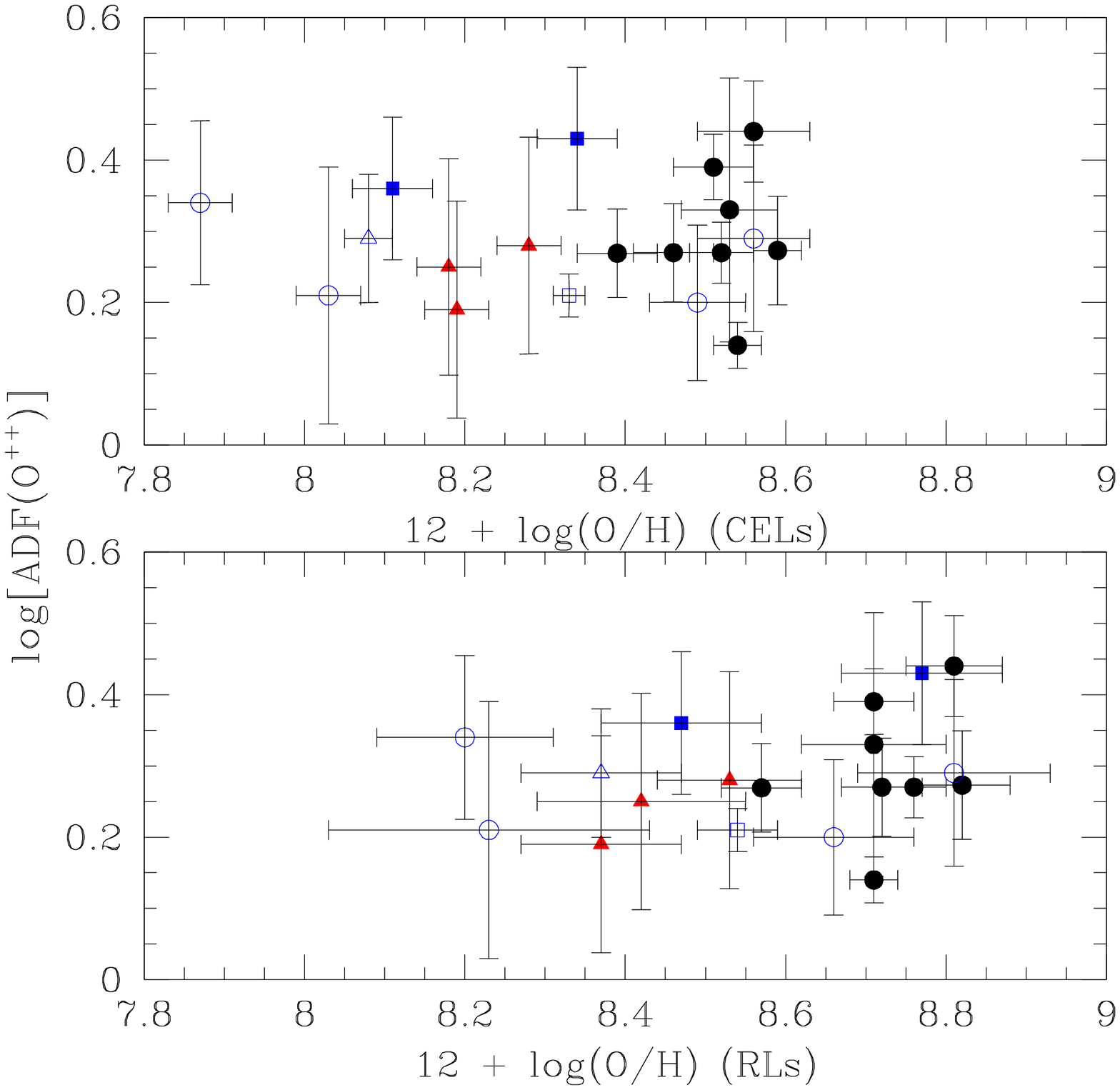}
\includegraphics[width=5cm,angle=0]{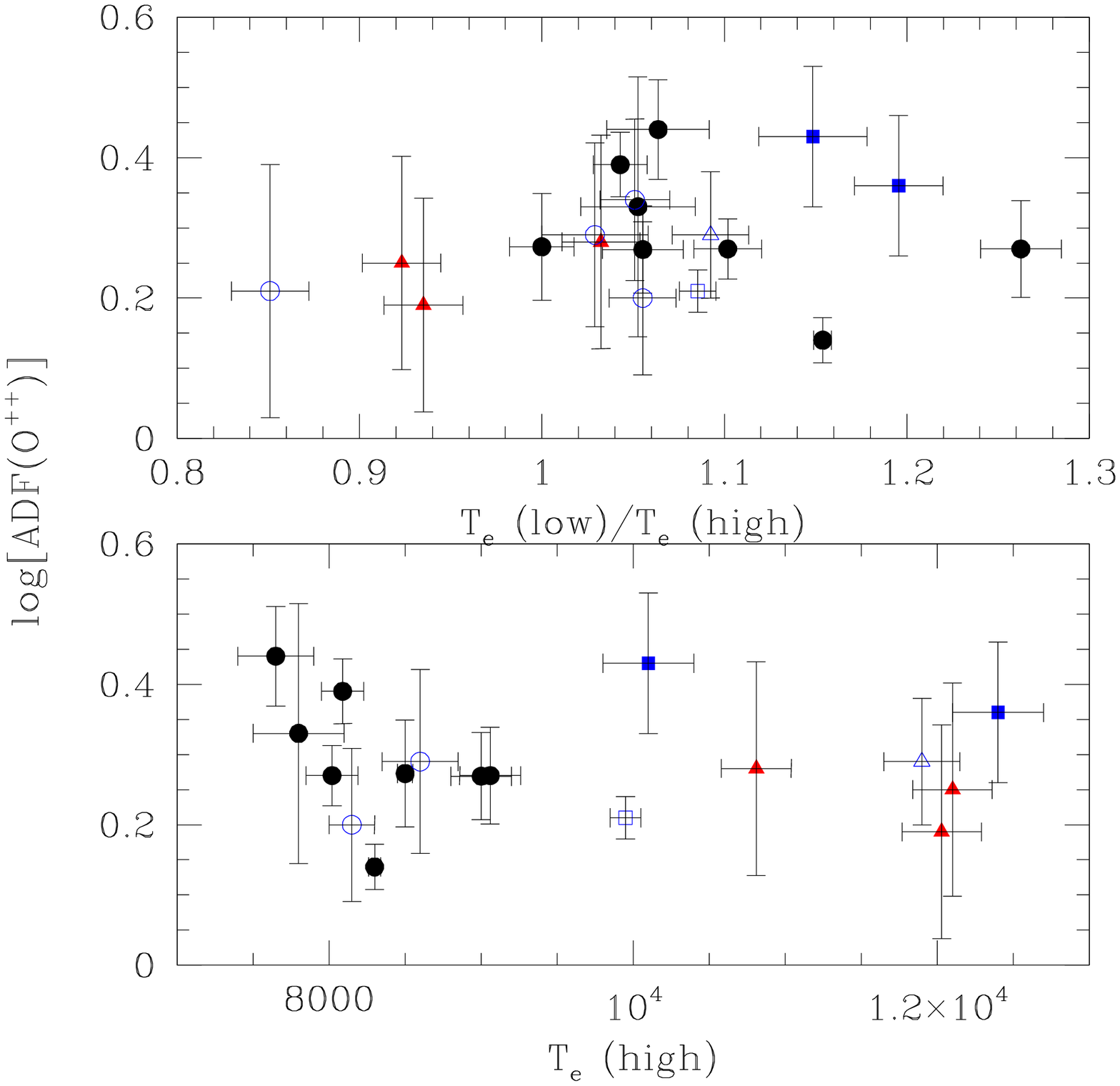}
\protect\caption[ ]{ADF(O$^{++}$) $vs.$ oxygen total abundances from CELs and RLs (left) and $vs.$ $T_e$(high) and 
$T_e$(low)/$T_e$(high). 
Symbols are the same than in Figure~\ref{adfoxy}.
\label{adf_Orl_cel}}
\end{figure}

On the other hand, several authors have found that the large range of ADFs found for PN is because the ADF 
increases monotonically with metallicity (see e.g. \cite{liuetal00}). 
In the case of H\thinspace {\sc ii} regions, this effect has not been found (see Figure~\ref{adf_Orl_cel}), at 
least in the studied metallicity range. We have also represented ADF $vs.$ $T_e$(high)/$T_e$(low) and we have found 
no correlation (see Figure~\ref{adf_Orl_cel}), indicating that the large scale variations of $T_e$ due to the 
natural gradients of this parameter along the nebulae seem not to be related with the ADF. 
Finally, we have also check that the ADF do not depend on the assumed temperature (see Figure~\ref{adf_Orl_cel}), 
which discards systematical effects in the abundance determination from CELs.

It is clear that it is necessary to increase the sample of H\thinspace {\sc ii} regions in which RLs have been 
measured with a good signal-to-noise. A systematic search of RLs in extragalctic H\thinspace {\sc ii} regions would 
permit us to cover a larger fraction of the nebular volume, to increase the metallicity range. 

Finally, \cite{stasinskaszczerba01} proposed that photoelectric heating due to dust grains could increase $T_e$ in 
the zones near the central star of a PN. 
\cite{robertsontessigarnett05} found that the ADF was not correlated with the effective temperature of the ionizing 
star for a sample of PN. 
For our sample of H\thinspace {\sc ii} regions we have not found a correlation between the ADF and the effective 
temperature of the ionizing star --or spectral type, which could give an idea of the radiation hardness-- of each 
nebulae (see \cite{garciarojas06}).

\section{Implications for the temperature fluctuations scenario.\label{discusiont2}}

Many of the arguments against temperature fluctuations are based on that photoionization models can not reproduce 
observed $t^2$'s, but photoionization models could not be enough realistic, and additional energy sources could be 
necessary to explain the observed discrepancies among observational data and photoionization models predictions  
(see \cite{viegas02} and references therein).

\begin{figure}[ht!]
\centering
\includegraphics[width=5cm,angle=0]{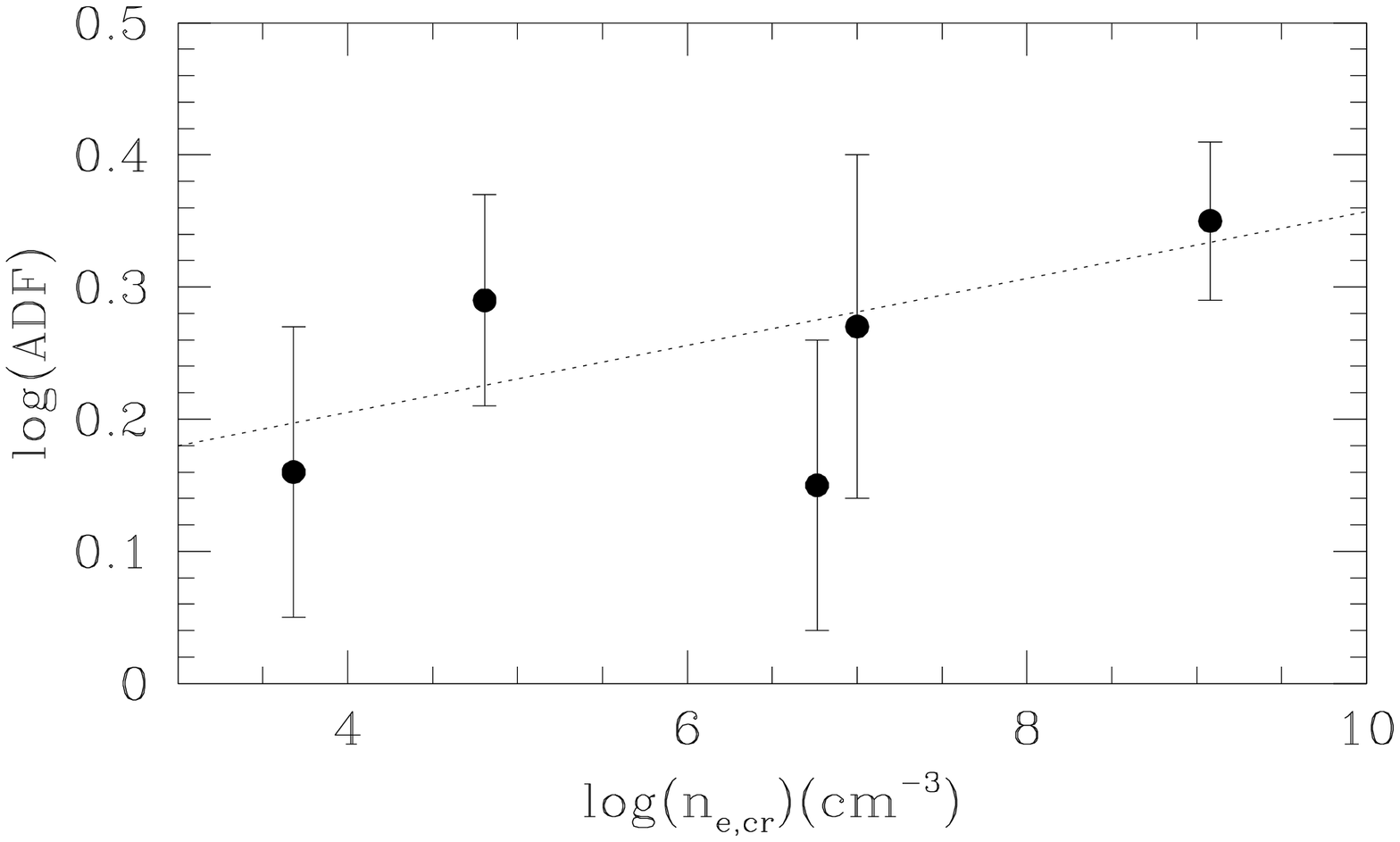}
\includegraphics[width=5cm,angle=0]{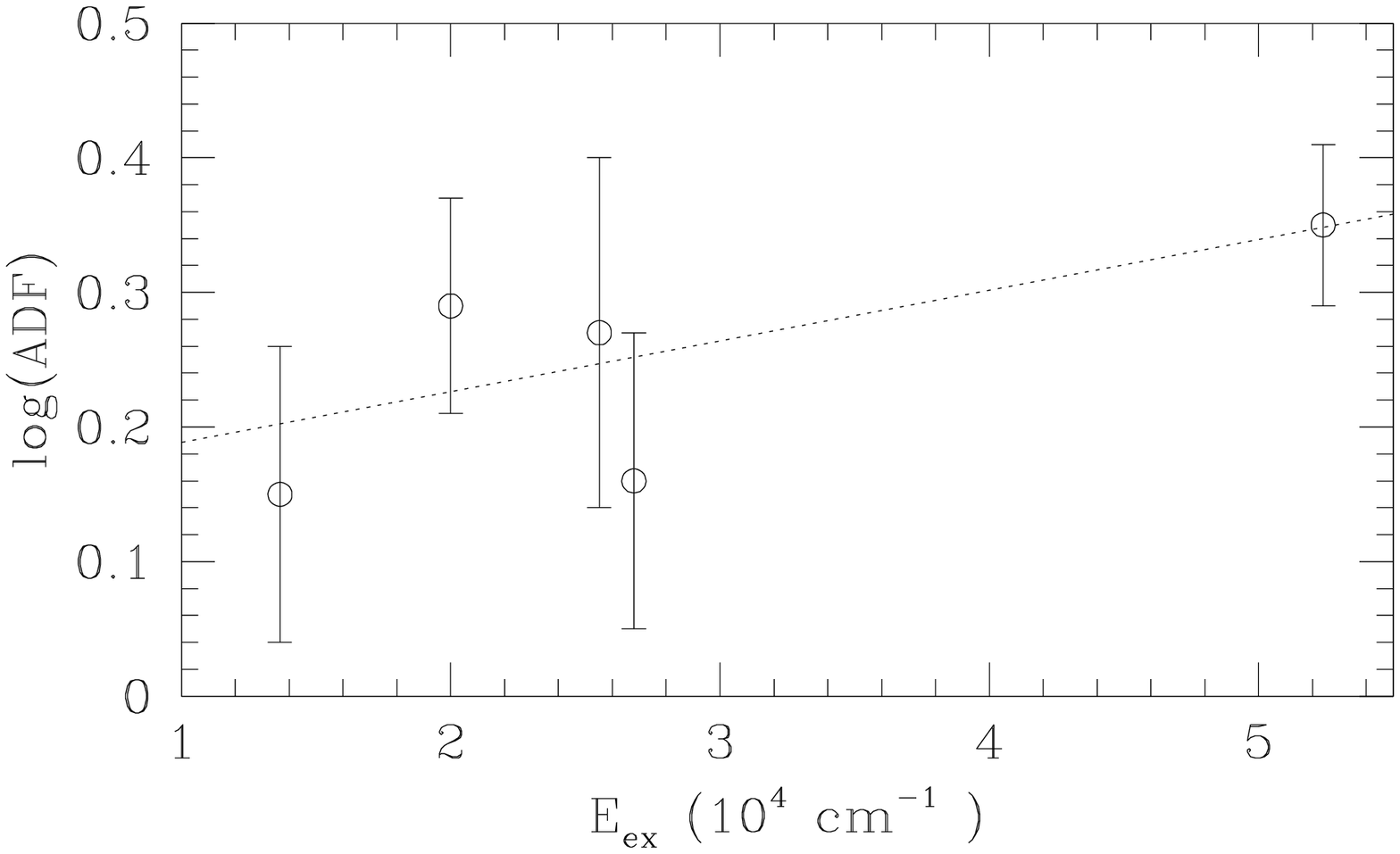}
\protect\caption[ ]{ADF for different ions $vs.$ critical density (left panel) and $vs.$ excitation energy of the 
upper level of the line (right panel).
\label{adf_exc_ncr}}
\end{figure}

In the last years, several observational evidences have been presented against temperature fluctuations in PN (see 
\cite{liu06} and references therein). On the other hand, very recently, \cite{lopezsanchezetal06} reported some 
reasons against the two-phase model of \cite{tsamispequignot05}. Here, we present additional objections to the 
two-phase model. 
By definition, under the temperature fluctuations scenario, the ADF should be related to the excitation energy (see 
\cite{peimbert67}); on the other hand, under the density fluctuations scenario (\cite{viegasclegg94}) the abundance 
discrepancy should be higher if the abundance has been computed from a CEL with an upper level with low critical 
density, $n_{e, cr}$, so ADF and critical density should be inversely proportional. 
\cite{liuetal00} and \cite{liuetal01} compared several abundance determinations in PNe (CELs in UV, optical and 
far-IR), and showed that the ADF was not related with the excitation energy, nor with $n_e$$_{, cr}$. 
As it has been pointed out in previous works (see \cite{garciarojasetal06, garciarojasetal06b}), we can not compare 
abundances derived from CELs in different ranges (UV, optical or far-IR) of the same ion, because UV and far-IR 
observations covering the same slit areas in the optical range are not available. Nevertheless, we can compare the 
ADFs obtained for Galactic and extragalactic H\thinspace {\sc ii} regions, with the excitation energy and with the 
critical density, $n_e$$_{, cr}$, of the upper level of the main CEL of each ion. 
In Figure~\ref{adf_exc_ncr} we shown a possible correlation (r=0.6) between the ADF and the excitation energy, 
E$_{ex}$; there is also an apparent correlation (r=0.5) with $n_e$$_{, cr}$, contrary to the predictions of the 
density fluctuations theory. 
These results are an additional probe that the abundance discrepancy scenario in H\thinspace {\sc ii} regions 
should be different to those proposed for PNe, so results on PNe can not be extrapolated to H\thinspace {\sc ii} 
regions. 

\subsection{Temperatures from O\thinspace {\sc ii} RLs.\label{te_rls}}

The chemically inhomogeneous or two phase model (\cite{liuetal01}, \cite{tsamisetal03} and 
\cite{tsamispequignot05}) predicts  
$T_e$(RLs) $\le$ $T_e$(He\thinspace{\sc i}) $\le$ $T_e$(H\thinspace{\sc i}) $\le$ $T_e$(CELs) 
(\cite{liu03}), with the difference between temperatures being proportional to ADF (see e.g. Figure~8 of 
\cite{liuetal01}).

\cite{wessonetal03} used --for the first time-- the temperature dependent ratio 
$I$($\lambda$4089.29)/ $I$($\lambda$4649.14) 
to derive $T_e$ of the ionized gas from which O\thinspace {\sc ii} lines arise in PNe ($T_e$(RL)). 
These authors found very low $T_e$ in two H-deficient knots of the PN Abell 30. 
Later, in subsequent papers they have found similar results in other PNe (e.g. \cite{tsamisetal04, wessonetal05}). 
This method present several difficulties that were pointed out by \cite{tsamisetal04}.

\begin{figure}[]
\centering
\includegraphics[width=5.2cm,angle=0]{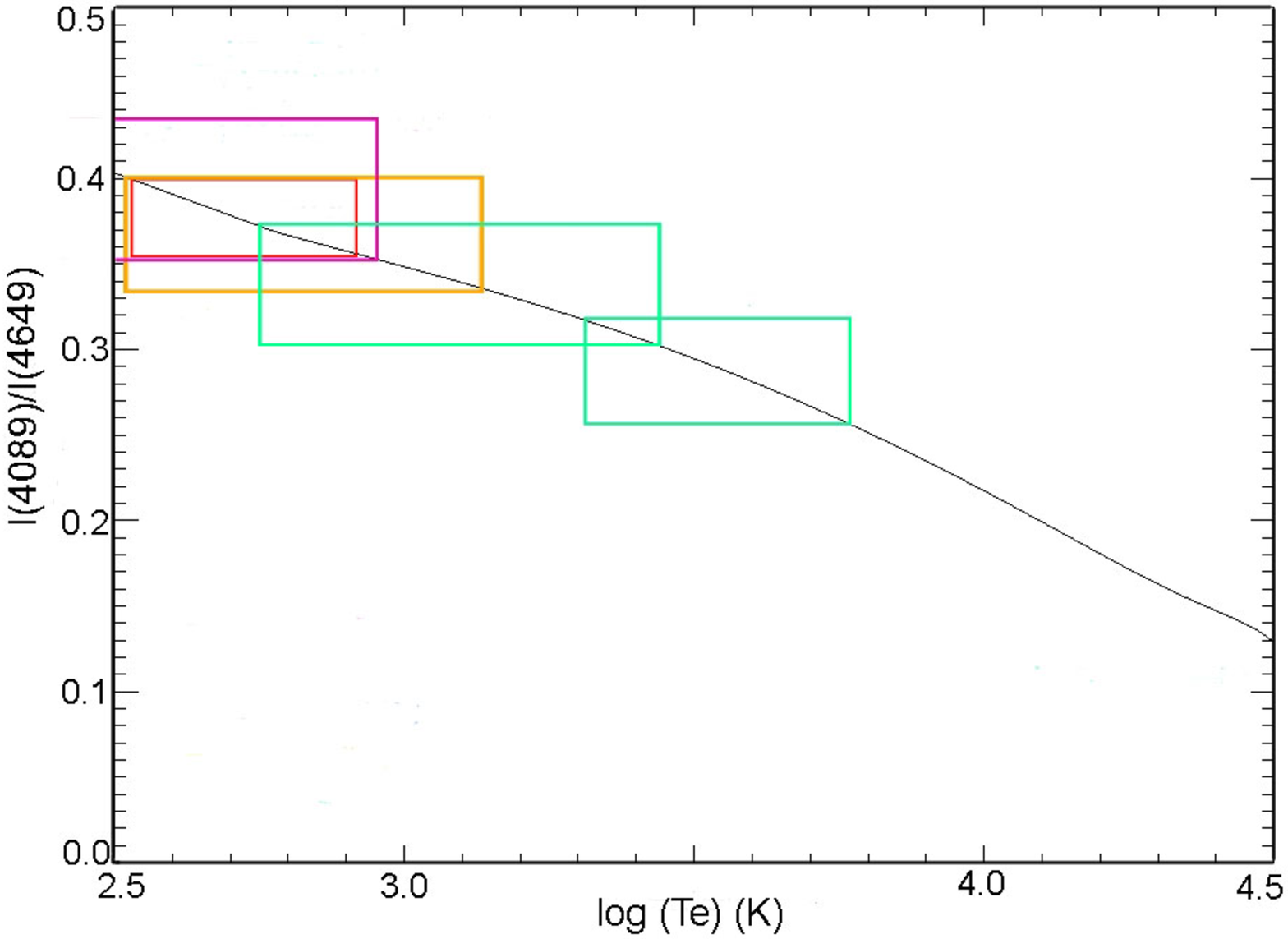}
\includegraphics[width=5.2cm,angle=0]{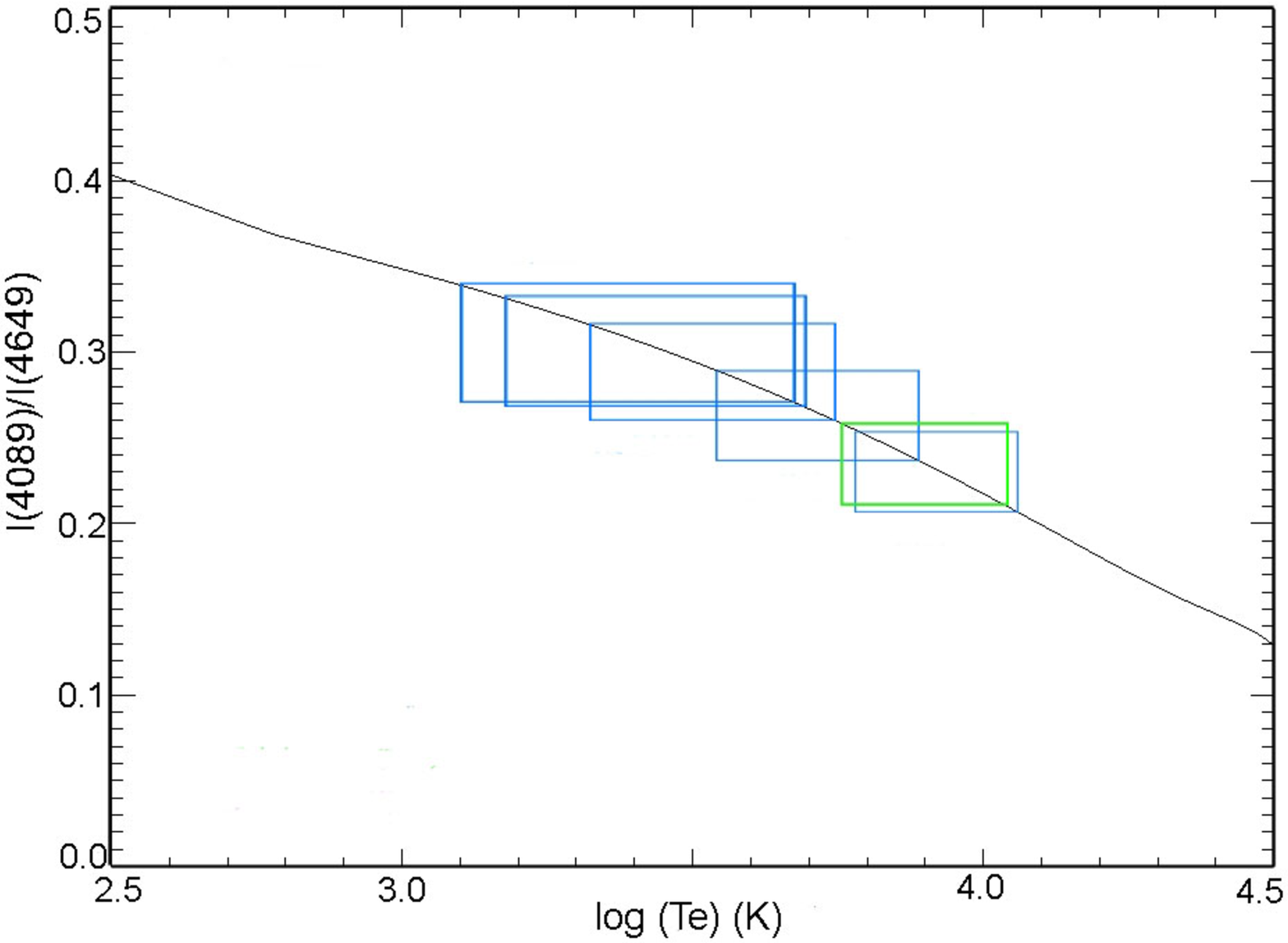}
\includegraphics[width=5.2cm,angle=0]{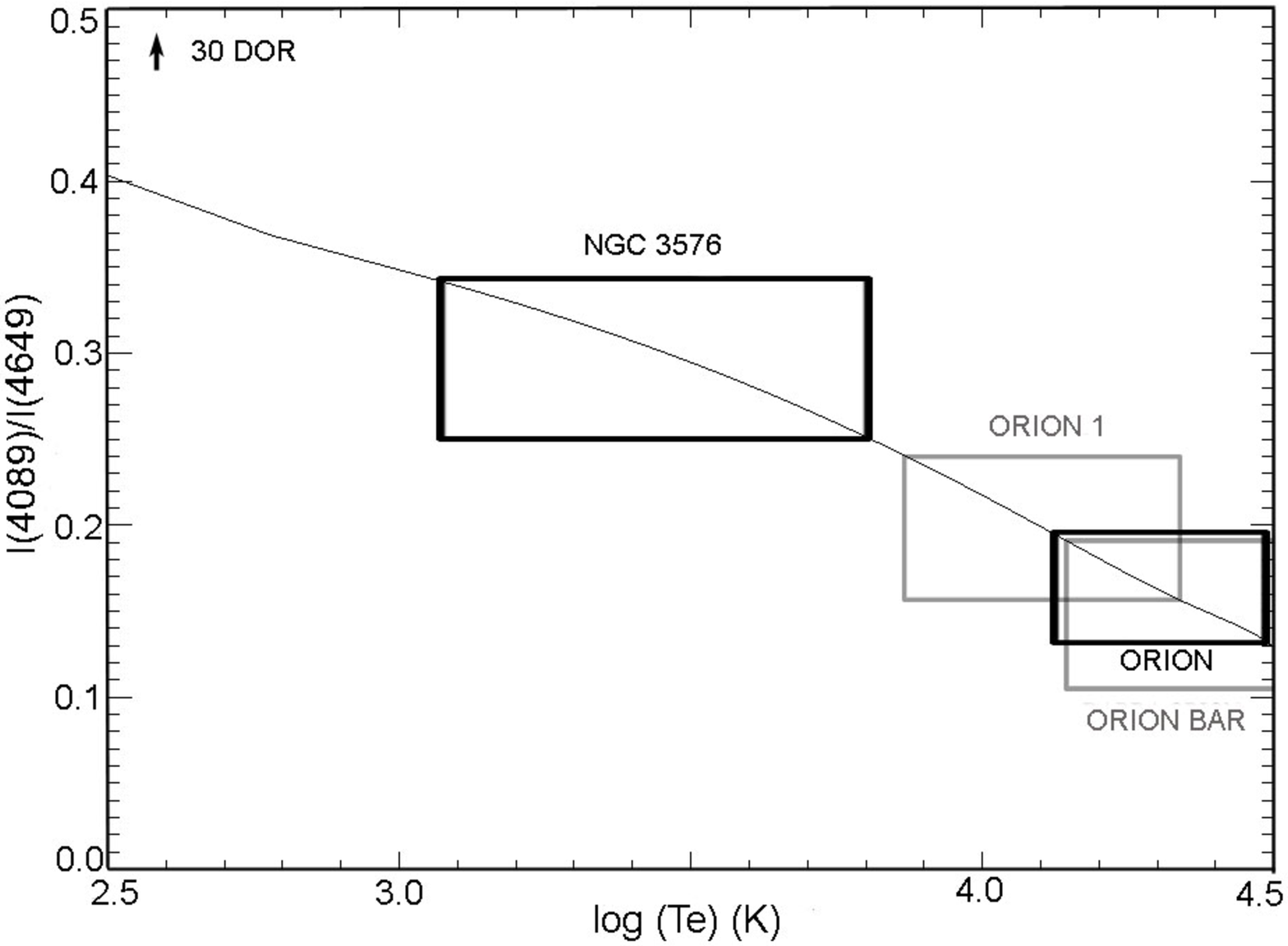}
\protect\caption[ ]{O\thinspace {\sc ii} $\lambda$4089.29/$\lambda$4649.14 ratio as a function of electron 
temperature. Solid line in the three panels represents the theoretical ratio. Observations are shown as error 
boxes. Vertical arrow shows an object with a ratio outside the scale. Panels are described in text. 
\label{temp_rls}}
\end{figure}

The determination of $T_e$ from O\thinspace {\sc ii} RLs in H\thinspace {\sc ii} regions is very difficult because 
the lines are usually much fainter than in bright PNe.  
In Figure~\ref{temp_rls} we have represented the dependence of the 
$I$($\lambda$4089.29)/$I$($\lambda$4649.14) ratio with $T_e$, and we have superimposed the ratios obtained in 
NGC~3576 and the Orion Nebula, the only objects in our sample in which it has been possible to detect and measure 
the O\thinspace {\sc ii} $\lambda$4089.29 line. 
Moreover, we have represented the line ratios measured in the giant H\thinspace {\sc ii} region 30 Doradus 
\cite{apeimbert03} and in several PNe from the literature: (most of them from the sample of \cite{tsamisetal04} and 
a sub-sample with ADF $>$4 from \cite{wessonetal05}).

As it is shown in the upper-left panel of Figure~\ref{temp_rls}, PNe with large ADFs show, in general, very low 
$T_e$(O\thinspace {\sc ii}), which is consistent with the two-phase model. On the other hand, PN with moderate 
values of the ADF (upper-right panel of Figure~\ref{temp_rls}) show $T_e$(O\thinspace {\sc ii}) that are, in most 
of the cases, smaller than $T_e$(CELs) and $T_e$([H\thinspace {\sc i}]). 
In the case of Galactic H\thinspace {\sc ii} regions, the temperature obtained for the Orion Nebula is somewhat 
larger than that obtained from CELs; the ratios obtained for NGC~3576 and 30 Doradus are higher, but  the 
O\thinspace {\sc ii} $\lambda$4089.29 line in both nebulae is affected by charge transfer effects in the CCD (due 
to the presence of very bright emission lines in adjacent orders) that can not be subtracted due to their low 
signal-to-noise ratio (see \cite{garciarojas06}), 
With the aim of obtain additional data for H\thinspace {\sc ii} regions, we have represented the ratios for two 
additional positions in the Orion nebula: the position labelled by \cite{estebanetal98} as ``Orion 1'', and a slit 
position placed on the Orion bar (24$''$ N and 12$''$ O from $\theta$$^2$Ori A, labelled as ``Orion Bar''). 
It can be see that $T_e$ values for this regions are similar to those obtained from the values obtained by  
\cite{estebanetal04}. 

\section{Conclusions.\label{conclu}}

We have presented here the results from deep high resolution spectrophotometry of H\thinspace {\sc ii} regions. 
From the analysis of these data, it is clear that the abundance discrepancies in H\thinspace {\sc ii} regions do 
not show the same behavior than in PNe, and could be explained assuming temperature fluctuations over the observed 
volume of the nebula. Moreover, the results obtained for PNe can not be extrapolated for H\thinspace {\sc ii} 
regions; in fact, it seems that the two-phase model scenario proposed by \cite{tsamispequignot05} is not suitable 
to explain the abundance discrepancy in H\thinspace {\sc ii} regions. 

%
%

%
%



\printindex
\end{document}